# The water – carbon monoxide dimer: new infrared spectra, *ab initio* rovibrational energy level calculations, and an interesting intermolecular mode

A. Barclay,[1] A. van der Avoird,[2] A.R.W. McKellar,[3] and N. Moazzen-Ahmadi[1]

[1] *Department of Physics and Astronomy, University of Calgary, 2500 University Drive North West, Calgary, Alberta T2N 1N4, Canada*

[2] *Theoretical Chemistry, Institute for Molecules and Materials, Radboud University, Heyendaalseweg 135,6525 AJ Nijmegen, The Netherlands*

[3] *National Research Council of Canada, Ottawa, Ontario K1A 0R6, Canada*

**Abstract**

Bound state rovibrational energy calculations using a high-level intermolecular potential surface are reported for $H_2O$-CO and $D_2O$-CO. They predict the ground $K = 1$ levels to lie about 20 (12) cm$^{-1}$ above $K = 0$ for $H_2O$-CO ($D_2O$-CO) in good agreement with past experiments. But the first excited $K = 1$ levels are predicted to lie about 3 cm$^{-1}$ *below* their $K = 0$ counterparts in both cases. Line strength calculations also indicate that mid-infrared transitions from the $K = 0$ ground state to this seemingly anomalous excited $K = 1$ state should be observable. These predictions are strikingly verified by new spectroscopic measurements covering the C-O stretch region around 2200 cm$^{-1}$ for $H_2O$-CO, $D_2O$-CO, and HOD-CO, and the O-D stretch region around 2700 cm$^{-1}$ for $D_2O$-CO, HOD-CO, and DOH-CO. The experiments probe a pulsed supersonic slit jet expansion using tunable infrared quantum cascade laser or optical parametric oscillator sources. Discrete perturbations in the O-D stretch region give an experimental lower limit to the binding energy $D_0$ of about 340 cm$^{-1}$ for $D_2O$-CO, as compared to our calculated value of 368 cm$^{-1}$. Wavefunction plots are presented to help understand the intermolecular dynamics of $H_2O$-CO. Coriolis interactions are invoked to explain the seemingly anomalous energies of the first excited $K = 1$ levels.

## I.  Introduction

The weakly-bound $H_2O$-CO dimer has a planar equilibrium structure with the heavy atoms (O, C, O) in an approximately collinear configuration and a hydrogen bond between the water and the carbon of CO. Proton tunneling, which interchanges the water H atoms, gives rise to two easily resolved tunneling states which correspond to distinct nuclear spin modifications and thus correlate with *para*- and *ortho*-water. The magnitude of the resulting splitting in the ground rotational state is about 0.9 cm$^{-1}$ for $H_2O$-CO, or about 0.04 cm$^{-1}$ for $D_2O$-CO. For the mixed isotope species containing HDO, there is of course no tunneling because the H and D atoms are distinguishable. Instead, there are two isomers which we can call HOD-CO (deuteron bound) and DOH-CO (proton bound). The former species has a lower zero-point energy and is thus easier to observe (the energy difference is estimated to be about 12 cm$^{-1}$).[1] Due to the almost linear heavy atom configuration, $H_2O$-CO has a large $A$ rotational constant, equal to about 19 cm$^{-1}$ (or 12 cm$^{-1}$ for $D_2O$-CO), which means that the $K_a$ quantum number is highly significant (denoted $K$ in the remainder of this paper). High resolution spectroscopic studies of water-CO were first made by Yaron et al.[2], who measured pure rotational microwave transitions within the $K = 0$ manifold, and by Bumgarner et al.[3], who measured millimeter wave transitions with $K = 1$ ← 0. So far, no spectra involving higher $K$ values have been detected. There has been considerable progress in observing mid-infrared rotation-vibration spectra. Thus Brookes and McKellar[1] studied the C-O stretch region (4.7 μm), Oudjeans and Miller[4] the O-H stretch region (2.7 μm), Zhu et al.[5] the $D_2O$ bending region (8.5 μm), and Rivera-Rivera et al.[6] the $H_2O$ bending region (6.3 μm).

There have been many quantum chemical investigations of water-CO intermolecular interaction,[7-11] but to our knowledge only a few high-level *ab initio* calculations of the global? potential energy surface. Wheatley and Harvey[12] computed a seven-dimensional surface (including CO stretching and water bending) using extrapolated intermolecular perturbation theory, and used it to determine second virial coefficients. More recently, Rivera-Rivera et al.[6] reported various five- and six-dimensional potentials, beginning with *ab initio* calculations [CCSD(T)/aug-cc-pVTZ, or MP2/aug-cc-pVNZ] and then "morphing" for better agreement with experiment. Finally, Kalugina et al.[13] computed a five-dimensional potential at a higher level of theory [CCSD(T)-F12a] and used it to calculate rotational excitation cross-sections for astrophysical applications.

In the theoretical part of the present paper, we use the latter potential surface[13] to make detailed calculations of rotational and vibration-rotation energy levels for $H_2O$-CO and $D_2O$-CO. The results agree well with experiment, and are especially useful for understanding a newly observed intermolecular vibration mode. In the experimental part of the paper, we study $D_2O$-CO, HOD-CO, and DOH-CO spectra in the O-D stretch region (3.6 µm) for the first time and extend the previous C-O stretch region results on $H_2O$-CO, $D_2O$-CO, and HOD-CO to include the $K = 1 \leftarrow 0$ bands. More significantly, in both regions we also observe combination bands involving the intermolecular mode just mentioned, which is the in-plane CO bend. This mode, which has $K = 1$ and is observed for $H_2O$-CO, $D_2O$-CO, and HOD-CO, is particularly interesting. For one thing, it exhibits anomalous isotope shifts for example, the $D_2O$-CO combination state is only slightly lower in energy than the $H_2O$-CO state even though their $A$-values are quite different (as mentioned above). Although we do not observe the $K = 0$ level of the new mode, our calculations, which are well supported by experiment, show that it lies *above* the $K = 1$ level. While similar inversion of the normal ordering of $K$ states has been observed in other weakly-bound complexes, it is especially notable here because $H_2O$-CO is a simple and relatively rigid species (except of course for the proton tunneling) in which hindered internal rotation plays no role.

Since it has a planar equilibrium structure, vibrational modes of water-CO can be classified as either in-plane or out-of-plane, giving symmetry species $A'$ or $A''$, respectively, in the $C_s$ point group. The four possible intramolecular fundamental vibrations are all $A'$ (C-O stretch, $H_2O$ bend, symmetric O-H stretch, antisymmetric O-H stretch). The five possible intermolecular fundamentals are either $A'$ (in-plane CO bend, van der Waals stretch, in-plane water bend) or $A''$ (out-of-plane CO bend, out-of-plane water bend). The two tunneling components correspond to the two possible relative spin orientations of the H or D nuclei, aligned or anti-aligned, and in previous spectroscopic studies they have been labeled as the A and B components.[1-6] The ground state, A, has relative statistical weights 1 ($H_2O$-CO) or 2 ($D_2O$-CO) while the B state has weights 3 or 1. These tunneling states correspond to the nuclear spin modifications of the isolated water molecule: A correlates with *para*-$H_2O$ and *ortho*-$D_2O$, and B with *ortho*-$H_2O$ and *para*-$D_2O$.

## II.  Theoretical rovibrational energies and line strengths

The potential energy surface of Kalugina et al.[13] used here has a global minimum with depth $D_e = 646.1$ cm$^{-1}$, an intermolecular center-of-mass separation $R$ of 3.93 Å and a planar configuration with $\theta_1 = 115.3°$ and $\theta_2 = 106.7°$. The angles $\theta_1$ and $\theta_2$ were defined in the coordinate system shown in Fig. 1 of Ref. 13. Here, we use different angular coordinates: a set of 4 Euler angles that define the orientations of the water and CO monomers with respect to a dimer-fixed frame with its z-axis along the vector $\mathbf{R}$ that points from the center of mass of water to the center of mass of CO, see Fig. 1. In this frame $\theta_1$ is the angle between the twofold symmetry axis of $H_2O$ (or $D_2O$) and the vector **R**, and $\theta_2$ is the angle between the CO axis and the vector **R**. The angle $\phi$ is the dihedral (or torsional) angle between the plane through **R** and the $H_2O$ symmetry axis and the plane through **R** and the CO axis, and $\chi$ is the angle of rotation of $H_2O$ or $D_2O$ around its symmetry axis. The equilibrium values are $\theta_1 = 115.3°$, $\theta_2 = 8.6°$, $\phi = 0°$, and $\chi = 0°$ for $H_2O$-CO. The equilibrium angles $\theta_1 = 115.0°$ and $\theta_2 = 8.3°$ are slightly different for $D_2O$-CO, because of the shift of the center of mass of $D_2O$ with respect to $H_2O$. The angles $\varphi$ and $\chi$ being zero implies that the equilibrium structure of the dimer is planar, the small value of $\theta_2$ indicates that the CO bond axis is nearly parallel to the intermolecular axis **R**. This OH…C hydrogen bonded structure agrees well with experiment.[2,3] There is also an OH…O-bonded local minimum with a depth of 340.4 cm$^{-1}$ and a separation $R$ of 3.70 Å. For comparison, some previous *ab initio* global minimum energies were $D_e = 629$,[12] 622,[11] and 650 cm$^{-1}$.[6]

Bound state $H_2O$-CO rovibrational levels with $J = 0$ and 1 were calculated in Ref. 13, and these calculations are extended here with $J = 2$ levels. Furthermore, the $H_2O$-CO potential surface was re-expressed in the dimer-fixed coordinates of $D_2O$-CO and also the rovibrational levels of $D_2O$-CO were calculated for $J = 0$, 1 and 2. The method[13,14] is based on a computational approach[15] developed for weakly bound molecular dimers with large amplitude internal motions, and is similar to a close-coupling scattering calculation. The $H_2O$ (or $D_2O$) and CO monomers were considered to be rigid and we used ground state experimental values for the rotational constants of $H_2O$ ($A = 27.8806$ cm$^{-1}$, $B = 9.2778$ cm$^{-1}$, $C = 14.5216$ cm$^{-1}$), $D_2O$ ($A = 15.4200$ cm$^{-1}$, $B = 4.8453$ cm$^{-1}$, $C = 7.2730$ cm$^{-1}$), and CO ($B = 1.9225$ cm$^{-1}$). Atomic masses are 1.007825 u for H, 2.014102 u for D, 12 u for $^{12}$C, and 15.994915 u for $^{16}$O. The water rotational levels are labeled by angular momentum $j_1$ and its projections $k_a$ and $k_c$ along the axes of smallest and largest moments of inertia, respectively. The CO rotational energy levels are designated by $j_2$. The discrete variable representation grid for the intermolecular center-of-mass distance $R$ contained 164 equidistant points ranging from 4.5 to 20 $a_0$, and a radial basis of 20 functions was contracted as in Ref.14. The angular basis was truncated to $j_1 \leq 9$ and $j_2 \leq 14$ for $H_2O$-CO and to $j_1 \leq 11$ and $j_2 \leq 14$ for $D_2O$-CO; higher $j_1$ values were needed for the latter complex because the rotational constants of $D_2O$ are about half of the $H_2O$ values. Energies are thus converged to about 0.01 cm$^{-1}$, but the energy differences are much more accurate than this.

The calculations were simplified by using the permutation-inversion symmetry group $G_4$. The so-called feasible symmetry operations generating this group are the permutation $P_{12}$ that interchanges the two protons in $H_2O$-CO (or deuterons in $D_2O$-CO) and the overall inversion of the system, $E^*$. States that are even/odd with respect to $P_{12}$ correspond to the *para/ortho* nuclear spin species. With our dimer-fixed coordinates inversion corresponds to reflection in the plane of the equilibrium geometry, so that states with even or odd parity under $E^*$ have $A'$ or $A''$ symmetry.

Our calculated binding energies $D_0$ are 315.98 and 368.42 cm$^{-1}$ for *para*-$H_2O$-CO and *ortho*-$D_2O$-CO, respectively, which shows that the intermolecular rovibrational zero-point energy is almost half of the well depth $D_e$ for $D_2O$-CO and even larger for $H_2O$-CO. The corresponding ground state energies of *ortho*-$H_2O$-CO and *para*-$D_2O$-CO, -315.16 and -368.39 cm$^{-1}$, are very similar to those of *para*-$H_2O$-CO and *ortho*-$D_2O$-CO, which demonstrates already that the internal rotations of the monomers in these complexes are strongly hindered. The calculated ground state $K = 0$ tunneling splittings are 0.795 ($H_2O$-CO) and 0.030 ($D_2O$-CO) cm$^{-1}$. For comparison, Rivera-Rivera et al.[6] obtained binding energies $D_0$ around 338 cm$^{-1}$ for $H_2O$-CO and 391 cm$^{-1}$ for $D_2O$-CO. They also calculated ground state $K = 0$ tunneling splittings of 0.83 and 0.04 cm$^{-1}$, respectively.

Since transitions connecting the A and B states are quite strongly forbidden, there are no direct experimental determinations of the tunneling splittings. However, the sum of the absolute values of the splittings in the $K = 0$ and 1 states can be measured precisely, and the resulting experimental sums are 1.1130 ($H_2O$-CO) and 0.0676 cm$^{-1}$ ($D_2O$-CO).[3] In previous experimental papers, the $K = 0$ and 1 splittings were assumed to be equal (and opposite), giving values of 0.557 ($H_2O$-CO) and 0.034 ($D_2O$-CO) cm$^{-1}$. But as we now know, calculations give a strong $K$-dependence, for example from 0.795 ($K = 0$) to 0.203 cm$^{-1}$ ($K = 1$) for $H_2O$-CO. Our calculated sums of $K = 0$ and 1 splittings are 0.9975 and 0.0493 cm$^{-1}$, which underestimate the true values by about 10% ($H_2O$-CO) or 30% ($D_2O$-CO). This suggests the following scaled estimates for the true individual splittings: $H_2O$-CO, 0.875 ($K = 0$) and 0.238 cm$^{-1}$ ($K = 1$); $D_2O$-CO, 0.041 ($K = 0$) and 0.026 cm$^{-1}$ ($K = 1$).

The complete list of calculated levels is given as Supplementary Material. Organizing these levels into rotational stacks with energies equal to $BJ(J + 1)$ with $B \approx 0.092$ ($H_2O$-CO) or 0.087 cm$^{-1}$ ($D_2O$-CO) made it straightforward to determine $K$-values for almost every stack of levels, and the origins of the resulting $K$-stacks are shown in Table 1 for the A and B states of $H_2O$- and $D_2O$-CO, with the zero of energy taken as the lowest bound $J = 0$ level. (Note however the almost coincident $D_2O$-CO stacks with origins at about 89.2 cm$^{-1}$ where the $K = 1$ or 2 labeling was not quite straightforward.) The left-hand side of Table 1 gives vibrational mode assignments. These assignments are fairly clear for the lowest few $K = 0$ stacks, aided by the fact that even and odd parity distinguishes between $A'$ and $A''$ modes. Thus in Table 1 we have: ground state, in-plane CO bend, out-of-plane CO bend, van der Waals stretch, in-plane bend overtone, etc. However, for $K = 1$ and especially $K = 2$, the mode labels become more ambiguous, as there are large Coriolis interactions which mix vibrational states. Above about 100 cm$^{-1}$, even labeling $K = 0$ levels becomes difficult. However, we can say that the second calculated $K = 0$ states with negative parity, at 125 ($H_2O$-CO) or 114 cm$^{-1}$ ($D_2O$-CO), must be the out-of-plane $H_2O$ bend and/or the combination of the in- and out-of-plane CO bends. The in-plane water bend is of special interest since according to Bumgarner et al.[3] it corresponds to the tunneling coordinate. These authors estimated this bending fundamental to lie around 120 cm$^{-1}$ in $H_2O$-CO using a simple one-dimensional tunneling model. But there are a number (at least five?) of other $A'$ overtone and combination vibrations in the 100 to 150 cm$^{-1}$ range, so the in-plane water bend fundamental may well be 'contaminated' due to anharmonic mixing with these other modes.

With the calculated wavefunctions we also computed the line strengths of the transitions from the ground state to various intermolecular modes in combination with the intramolecular C-O or O-H (O-D) stretch states. Since it may be assumed that the monomer transition moments are only weakly affected by the noncovalent intermolecular interactions, we constructed a transition dipole function containing the appropriate monomer transition dipole moment vector, expressed as a function of the intermolecular coordinates in the complex. The C-O stretch dipole vector simply lies along the CO bond axis. The $D_2O$ monomer vibration excited in $D_2O$-CO is the antisymmetric O-D stretch mode with its transition dipole moment vector in the plane of $D_2O$ perpendicular to its twofold symmetry axis. Table 1 gives the calculated relative line strengths for $R(0)$ transitions in the CO stretch region. The O-H or O-D stretch line strengths (not shown) have similar orders of magnitude. These line strengths are for transitions originating in the ground $K = 0$ state (there is very little population in $K = 1$ or higher states in our experiments, where the effective rotational temperature is about 2.5 K). The strongest transition by far is of course the fundamental $K = 0 \leftarrow 0$ parallel ($a$-type) band, while the fundamental $K = 1 \leftarrow 0$ perpendicular (*b*-type) band is at least 20 times weaker. This is as expected, since the CO monomer is closely aligned with the *a*-axis of the complex. Interestingly, the line strength of the $K = 1 \leftarrow 0$ transition to the combination involving the CO in-plane bend is comparable in strength to the fundamental $K = 1 \leftarrow 0$ band, while the corresponding $K = 0 \leftarrow 0$ CO in-plane bend band is much weaker. As we will see below, this prediction is confirmed by experiment. The in-plane CO bend is anomalous in that $K = 0$ is calculated to lie *higher* in energy than $K = 1$ by about 3 cm$^{-1}$ for both H$_2$O-CO and D$_2$O-CO. All other possible combination bands involving intermolecular modes are calculated to be thousands of times weaker than the $K = 0 \leftarrow 0$ fundamental, except for the $K = 1 \leftarrow 0$ band of the out-of-plane CO bend which is still about 300 times weaker.

Rivera-Rivera et al.[6] have calculated intermolecular vibrational frequencies for H$_2$O-CO (their Table 13) which can be compared with our Table 1. Their first mode at 19.73 (A state) and 18.64 cm$^{-1}$ (B state) was labeled by them as $\nu_9$ "in-plane bend", but it clearly must be the ground state $K = 1$ level, which we calculate at 19.868/18.871 cm$^{-1}$ (relative to the respective $K = 0$ levels). Their second mode, labeled $\nu_8$ "out-of-plane bend", at 50.57/48.76 cm$^{-1}$ must really be the in-plane bend, which we calculate at 51.270/51.706 cm$^{-1}$ for $K = 0$ and 49.429/47.749 cm$^{-1}$ for $K = 1$. Their third mode, labeled $\nu_4$ "intermolecular stretch" lies at 78.49/78.76 cm$^{-1}$ and agrees very well with our van der Waals stretch at 78.807/78.278 cm$^{-1}$. It seems that they missed the out-of-plane bend which we calculate at 70.704/68.203 cm$^{-1}$. (Note also that in Tables 4 and 6 of Ref. 6, the titles should say "$K = 1 \leftarrow 0$" and the A and B state column headings should be interchanged.)

Density functional theory (DFT) methods have also been used to calculate H$_2$O-CO intermolecular vibrations. For example, Lundell and Latajka[9] report various intermolecular harmonic frequencies ranging from 63 to 413 cm$^{-1}$, and anharmonic frequencies ranging from 8 to 690 cm$^{-1}$. However, it is difficult for us to see any value in these results.

## III.    Experimental spectra

Spectra were recorded at the University of Calgary as described previously[16-18] using a pulsed supersonic slit jet expansion probed by a rapid-scan tunable infrared source. In the 4.7 μm region, the source was a quantum cascade laser, and in the 3.6 μm region, it was an optical parametric oscillator. The usual expansion mixtures contained about 0.01% $H_2O$ or $D_2O$ plus 0.02 − 0.06% CO in helium carrier gas with a backing pressure of about 10 atmospheres. Wavenumber calibration was carried out by simultaneously recording signals from a fixed etalon and a reference gas cell containing $N_2O$ or $C_2H_2$. Spectral assignment and simulation were made using the PGOPHER software.[19]

We use the following empirical rotational energy expression to fit the experimental spectra,

$$E = \sigma + B_{av}[J(J+1) - K^2] - D_J[J(J+1) - K^2]^2 \pm [(B - C)/4][J(J+1)]. \qquad (1)$$

Here, $K = K_a$, and $B_{av} = (B + C)/2$. Each vibrational state and $K$-value has its own origin, σ, and rotational parameters, $B_{av}$ and $D_J$. The parameter $(B - C)$ is needed only for $K = 1$ states. States with $K > 1$ have not been observed for water-CO in this or past studies. We give each $K = 1$ state its own origin, rather than using the $A$ rotational parameter, because (as shown above) $A$ is not well-defined for many excited intermolecular vibrations. The theoretical $K$= 1 and 2 origins shown above in Table 1 are consistent with this expression. In our spectroscopic analyses below, ground state parameters for $H_2O$-CO and $D_2O$-CO were fixed at the values determined by Bumgarner et al.,[3] after translating from their system (using $A$) to ours (separate parameters for $K$

= 0 and 1). Ground state parameters for HOD-CO were taken from Yaron et al.,[2] while those for DOH-CO were fitted to the present infrared data.

## A. $H_2O$ - CO

In the C-O stretch region, the $K = 0 \leftarrow 0$ fundamental band (2154 cm$^{-1}$) of $H_2O$-CO has been studied previously.[1] Here we also observe new $K = 1 \leftarrow 0$ bands at about 2173 and 2200 cm$^{-1}$. All three bands have two components, arising from the tunneling states labeled A and B with nuclear spin weights of 1 and 3, respectively. The new $K = 1 \leftarrow 0$ bands are illustrated on the left-hand side of Fig. 2, and fitted parameters are listed in Table 2. For consistency, we include here new parameters for the $K = 0 \leftarrow 0$ fundamental band, which are in good agreement with the previous data.[1]

The first $K = 1 \leftarrow 0$ band at 2172.2 and 2173.3 cm$^{-1}$ is the expected perpendicular component of the C-O stretch fundamental, analog of the ground state band observed by Bumgarner et al.[3] at 561 and 595 GHz. The present energies of 19.741 or 18.594 cm$^{-1}$ (A or B state), relative to the $K = 0$ fundamental, are about 0.5% smaller than those in the ground state, 19.834 or 18.721 cm$^{-1}$ (see Table 3). But what about the second $K = 1 \leftarrow 0$ band near 2200 cm$^{-1}$, which represents energies of 47.855 or 45.790 cm$^{-1}$ (A or B) relative to the $K = 0$ origins? This is obviously the predicted (Table 1) *combination* band involving the sum of the intramolecular C-O stretch and intermolecular in-plane CO bend modes. We were not able to detect the corresponding combination band with $K = 0 \leftarrow 0$, which is understandable in view of its predicted line strength (Table 1) which is 44 times weaker.

**B. D₂O - CO**

In the C-O stretch region, we observed $D_2O$-CO fundamentals with $K = 0 \leftarrow 0$ and $1 \leftarrow 0$, plus the combination band with $K = 1 \leftarrow 0$, the same three bands for as for $H_2O$-CO. Here the A and B tunneling components are more closely spaced and have nuclear spin weights of 6 and 3, respectively. Spectra are shown on the right hand side of Fig. 2. The $K = 0 \leftarrow 0$ band had been studied previously,[1] while the $K = 1 \leftarrow 0$ bands are new. But the energy of the fundamental $K = 1$ state was actually already known for the A tunneling component by means of the weak $K = 1 \leftarrow 1$ band. The current result confirms this earlier $K = 1$ analysis,[1] extends it to the B component, and provides more accurate $K = 1$ parameters.

In the O-D stretch region, there are two possible fundamentals, corresponding to the symmetric ($\nu_1$, 2671.645 cm⁻¹) and antisymmetric ($\nu_3$, 2787.718 cm⁻¹) vibrations of $D_2O$ monomer.[20] In $D_2O$-CO we can also think of these vibrations in terms of free and bound O-D stretches, where the bound stretch involves the D atom participating in the bond with the C atom of the CO. The true situation is likely a mixture of these two pictures. We label the vibrations here as O-D stretch-1 (symmetric) and O-D stretch-3 (antisymmetric). We first detected the O-D stretch-3 $K = 0 \leftarrow 0$ fundamental near 2781 cm⁻¹, as well as the corresponding fundamental and combination $K = 1 \leftarrow 0$ bands, as shown in Fig. 3. We subsequently detected the weaker O-D stretch-1 $K = 0 \leftarrow 0$ and $1 \leftarrow 0$ fundamentals near 2665 and 2677 cm⁻¹, respectively, as shown in Fig. 4. Note that Oudjeans and Miller[4] only detected the antisymmetric $K = 0 \leftarrow 0$ fundamental in their study of $H_2O$-CO in the O-H stretch region. Our O-D stretch-1 bands are heavily perturbed, whereas the O-D stretch-3 bands are well-behaved. The perturbations are discrete, with weaker but still sharp perturbing lines appearing as satellites around the perturbed ones, giving matching patterns in the $P$- and $R$-branches for each upper state $J$-value. It's a textbook example of a bright state embedded in a moderately dense manifold of dark states. The source of the perturbation is discussed below. Its effects in the $K = 0 \leftarrow 0$ band made it impossible to separately assign A and B tunneling components. (The ground state $B$-values for the A and B components are very similar, so A and B cannot be distinguished on the basis of combination differences.) To analyze the $K = 0 \leftarrow 0$ band, we fitted the "center of gravity" of each $P$- and $R$-branch line, and thus obtained what we hope is an approximation of the deperturbed average of A and B. Although the $K = 1 \leftarrow 0$ band was very weak, it did seem possible to assign A and B components in the $P$- and $R$-branches in spite of the perturbations. The $K = 1 \leftarrow 0$ $Q$-branches are obviously present around $2676.75 - 2676.84$ cm$^{-1}$ (see Fig. 4), but they are also perturbed and remain unassigned in detail.

The fitted parameters for $D_2O$-CO are collected in Table 4. The energies of the $D_2O$-CO $K = 1$ states relative to $K = 0$ are summarized and compared with $H_2O$-CO and with theory in Table 3. We note that the fundamental $K = 1$ state energies in the C-O stretch region are very similar to those in the ground state ($\approx 11.7$ cm$^{-1}$), while those for O-D stretch-3 are somewhat smaller ($\approx 11.4$ cm$^{-1}$). An interesting fact is that the $K = 1$ combination states for $D_2O$-CO ($\approx 44$ cm$^{-1}$) are only a bit lower in energy than those of $H_2O$-CO ($\approx 47$ cm$^{-1}$), a further apparent anomaly for this state which however agrees well with our calculations in Table 1. Normally, $D_2O$-CO would be lower by the difference in $A$-values ($\approx 7.5$ cm$^{-1}$) *plus* the difference in the vibrational mode frequency itself, which deuteration should reduce.

### C. HOD – CO

The only previous spectroscopic observations of HOD-CO involved microwave[2] and C-O stretch infrared transitions,[1] both limited to $K = 0$. Here, we repeat the observation of the C-O stretch $K = 0 \leftarrow 0$ band, and also detect the same two $K = 1 \leftarrow 0$ bands as seen for $H_2O$- and

$D_2O$-CO. In addition, we observe the fundamental $K = 0 \leftarrow 0$ band in the O-D stretch region, but

not the $K = 1 \leftarrow 0$ bands. Results are shown in Fig. 5 and Table 5. This represents the first

observation of $K = 1$ states for HOD-CO. The fundamental $K = 1$ state energy, at 18.966 cm$^{-1}$ for

the C-O stretch region, is close to those of $H_2O$-CO, rather than $D_2O$-CO, which is as expected

since the D atom in HOD-CO lies close to the $a$-inertial axis and so has little effect on the $A$

rotational constant. The combination mode $K = 1$ state, at 48.096 cm$^{-1}$ for the C-O stretch region,

is even higher than those of $H_2O$-CO, another sign of an isotopic anomaly for this mode.

There is a very weak series located close to the HOD-CO $K = 0 \leftarrow 0$ band in the O-D

stretch region which is only barely visible in Fig. 5. It can be assigned to the HOD-$^{13}$CO

isotopologue on the basis of the observed line spacing, which matches that expected from the parameters measured by Yaron et al.[2] for this species. The isotopic shift in the O-D stretch frequency is -0.055 cm$^{-1}$ (see Table 5). The reason this band was detected at all is probably because its lines are noticeably sharper than those of the main band, a sign of reduced predissociation broadening in HOD-$^{13}$CO.

### D.  DOH - CO

The only previous spectroscopic observation of DOH-CO was of the fundamental $K = 0$

$\leftarrow 0$ band in the CO region. From observed relative line strengths, Brookes and McKellar

estimated its zero-point energy to be $12.4 \pm 2.5$ cm$^{-1}$ higher than that of HOD-CO.[1] We observe

the same band here, and confirm the exact rotational numbering given in Ref. 1, which had been

slightly uncertain. We also observe (very weakly!) the fundamental $K = 0 \leftarrow 0$ band in the O-D

stretch region, as shown in the upper right-hand panel of Fig. 5. Since there are no microwave observations of DOH-CO, we determine the ground and excited state parameters in a combined fit of the bands in the C-O and O-D regions. The results in Table 5 agree well with Ref. 1 for the ground and excited C-O stretch states. So far, we do not have any measurement of the $K = 1$ state for DOH-CO, but we can guess that its energy is fairly similar to that of $D_2O$-CO ($\approx 12$ cm$^{-1}$), just as HOD-CO is similar to $H_2O$-CO (see Table 3).

## IV.    Discussion

### A.    Rotational constants

The changes in rotational constant with vibrational excitation (alpha values) are a significant aspect of the parameters in Tables 2, 4, and 5. These are mostly small, but there are still systematic trends. Excitation of the C-O stretch causes slight decreases ($\approx 0.00004$ cm$^{-1}$) in $(B + C)/2$, as already noted in Ref. 1. Excitation of O-D stretch-3 also causes decreases, but they are mostly even smaller in magnitude. Excitation of $K$ from 0 to 1 causes noticeable increases for both the ground and excited C-O stretch states, especially for $H_2O$-CO. And finally, excitation of the $K = 1$ combination band states causes more significant increases in $(B + C)/2$, especially for $D_2O$-CO in the excited O-D stretch-3 state ($\approx 0.0002$ cm$^{-1}$). Since increases in rotational constant correspond to shorter bond lengths, it is interesting to note that water and CO actually become more closely bound upon excitation of $K$ from 0 to 1, and especially when accompanied by excitation of the intermolecular CO in-plane bend mode.

### B.    Vibrational shifts

As discussed previously,[1] $H_2O$-CO and $D_2O$-CO undergo substantial vibrational blue shifts in the C-O stretch region (+10.3 and +11.2 cm$^{-1}$, respectively). As well, the HOD-CO shift is similar to that of $D_2O$-CO, and the DOH-CO shift similar to $H_2O$-CO. In the O-D stretch region, we now find somewhat smaller red shifts for the O-D stretch-3 and O-D stretch-1 vibrations (-7.1 and -6.4 cm$^{-1}$) relative to the free $D_2O$ values.[20] The fact that they remain close to free $D_2O$ suggests that the appropriate picture for these modes is closer to antisymmetric and symmetric O-D stretch, rather than free and bound stretch. For $H_2O$-CO, a similar red shift was observed for the O-H stretch-3 vibration (-8.7 cm$^{-1}$).[4] For HOD-CO and DOH-CO in the O-D region, we observe vibrational shifts of -14.20 and +1.72 cm$^{-1}$, respectively, relative to free HDO.[21] Here the situation is obviously well described as bound O-D stretch for HOD-CO and free O-D stretch for DOH-CO.

### C. Perturbations

We have seen in Sec. III.B. that the O-D stretch-1 bands of $D_2O$-CO (Fig. 4) show numerous discrete perturbations, in contrast to the apparently unperturbed O-D stretch-3 (Fig. 3) and C-O stretch (Fig. 2) bands. At first this is a bit puzzling, since the nearest perturbing monomer state for O-D stretch-1, which is $D_2O$ $2\nu_2$, is fairly distant more than 300 cm$^{-1}$ lower in energy), while the nearest perturber for O-D stretch-3, which is O-D stretch-1, is much closer (<120 cm$^{-1}$). (To cause discrete perturbations as observed here, the perturbing state must lie below the perturbed one.) Evidently, the effective coupling between $\nu_1$ and $2\nu_2$ in $D_2O$-CO is stronger than that between $\nu_3$ and $\nu_1$, which of course is not unreasonable since the latter pair have different symmetries in the monomer.

Interestingly, the perturbations enable us to determine an experimental lower limit for the binding energy of $D_2O$-CO. We observe discrete (sharp) perturbations in the $K = 1 \leftarrow 0$ O-D stretch-1 band around 2677 cm$^{-1}$ (Fig. 4), which in practice means that this energy must lie below the threshold for dissociation of $D_2O$-CO into $D_2O$ in its $2\nu_2$ state and CO in its ground state. This $2\nu_2$ dissociation limit[20] lies at 2336.899 cm$^{-1}$ with respect to dissociation into ground state $D_2O$ and CO, which in turn means that the $D_2O$-CO binding energy must be greater than 340

cm$^{-1}$ in the ground state. This experimental value lies comfortably below our calculated binding energy $D_0$ of 368.42 cm$^{-1}$, but it does serve to illustrate how high-resolution spectroscopy can give specific information on the binding energy of a weakly-bound complex, in favorable circumstances. (Note that if the perturbations were instead due to CO stretch excitation in D$_2$O-CO, the lower limit would be 534 cm$^{-1}$, showing why we can be almost certain that $2\nu_2$ is the perturber).

### D. Calculated wavefunctions

In order to gain a better understanding of the intermolecular vibrational dynamics of water-CO, we turn to calculated wavefunctions. These wavefunctions depend on the five intermolecular coordinates defined with respect to a dimer-fixed frame in Sec. II, and illustrated in Fig. 1. For clarity the polar angles $\theta_1$ and $\theta_2$ of the H$_2$O symmetry axis and the CO bond axis, respectively, are here called $\theta$(H$_2$O) and $\theta$(CO). The angle of rotation of the water monomer about its twofold symmetry axis $\chi$ is called $\chi$(H$_2$O) and the torsional angle is called $\phi$. The intermolecular distance $R$ is here given in atomic units $a_0$, $1a_0 = 0.529177$ Å. Each wavefunction is visualized by means of four contour plots representing different two-dimensional cuts.

Wavefunctions for the ground $J = K = 0$ levels of H$_2$O-CO are shown in Fig. 6. Note that they are almost identical for the A and B states (*para* and *ortho*), except for the phase in the $\theta$(H$_2$O) vs. $\chi$(H$_2$O) plots: A is symmetric around $\chi = 90°$ and B is antisymmetric. The contours in these plots are concentrated near $\chi$(H$_2$O) $= 0°$ and $180°$, showing that the CO monomer center of mass is localized in the plane of the H$_2$O monomer, and also the $\theta$(H$_2$O) vs. $\theta$(CO) plots show the wavefunctions to be rather well localized around the equilibrium geometry. But the plots of $\phi$ vs. $R$ and $\theta$(CO) show that the wavefunctions are completely delocalized in the torsional angle, in other words, that this torsional motion is virtually free and that the CO axis does not care much whether it points in or out of the plane of the $H_2O$. We scanned the potential surface to get a better understanding of this torsional motion and to determine the barrier. When $\phi$ changes from 0° (the planar equilibrium structure) the CO monomer aligns more and more parallel to the intermolecular axis. That is, the optimum value of $\theta(CO)$ decreases from about 8° at equilibrium to 4° when $\phi$ is increased to 45°, and to 0° when $\phi = 90°$. Since CO is then precisely parallel to the intermolecular axis, the torsional angle has lost its meaning, just as West and East lose meaning near the North Pole. At the same time there is an increase of the optimum $\theta(H_2O)$, from 115° at equilibrium ($\phi = 0°$), to 120° at $\phi = 45°$ and 122° at $\phi = 90°$, while $R$ increases slightly, from 7.42 to 7.46 $a_0$. The barrier at $\phi = 90°$, with $\theta(H_2O)$ and $R$ relaxed, is less than 12 cm$^{-1}$, and the zero-point energy lies far above this. As $\phi$ increases further from 90° the system moves back to the equilibrium structure at $\phi = 180°$, $\theta(CO) = -8°$ (which is equivalent to $\phi = 0°$, $\theta(CO) = 8°$). So this nearly free internal motion along $\phi$ is clearly not a regular torsional motion. Of course, this weak $\phi$ dependence of the potential energy is an almost inevitable consequence of the very small value of $\theta(CO)$ for all $\phi$ values.

Wavefunction plots for the lowest $H_2O$-CO $K = 1$ state (origin at 19.868 cm$^{-1}$) are not shown here because they are very similar to those for $K = 0$ in Fig. 6. One small difference is that the amplitude in the $R$ vs. $\phi$ plot is slightly more concentrated close to planarity ($\phi = 0$).

Plots for the first excited $K = 0$ and 1 states of $H_2O$-CO are shown in Fig. 7 for the A state (the B state, not shown, continues to be similar). The assignment of these states to the in-plane CO bend vibration is supported by the existence of a node in the $\theta(CO)$ coordinate. This node is slightly tilted in the $\theta(H_2O)$ direction and also depends on $\phi$, which shows that this intermolecular vibration is a concerted motion. However, the similarity between the $K = 0$ and 1 plots here is much less close than it is for the ground $K = 0$ and 1 states, showing that their vibrational characteristics are different, and helping to explain their anomalous relative energies.

The second excited $H_2O$-CO $K = 0$ state at 70.704 cm$^{-1}$ has negative parity, so we already know that it involves an out-of-plane vibration. The wavefunction plots shown at the top of Fig. 8 reflect this out-of-plane character by having nodal planes for $\phi = 0°$ (and hence $\phi$ is fixed at 90° for the θ(CO) vs θ(H$_2$O) and χ(H$_2$O) vs. θ(H$_2$O) plots). The third excited $H_2O$-CO $K = 0$ state at 77.807 cm$^{-1}$ is shown at the bottom of Fig. 8, and the strong nodal character in the $R$ vs. $\phi$ plot supports its assignment to the intermolecular stretching vibration. Looking closely at the θ(CO) vs θ(H$_2$O) wavefunction plots for this state at 77.807 cm$^{-1}$ and those for the in-plane CO bend state at 51.270 cm$^{-1}$, we see some evidence that these two vibrations are slightly mixed with each other.

The wavefunction plots for the third excited $K = 1$ state at 96.531 cm$^{-1}$ (not shown) support its assignment to the $K = 1$ version of the intermolecular stretching vibration. However, assignments of the remaining states we investigated up to about 120 cm$^{-1}$ are not really clear from their wavefunctions. This includes the $K = 1$ state at 80.551 cm$^{-1}$, whose wavefunctions do not match very well with the $K = 0$ out-of-plane CO bend state at 70.704 cm$^{-1}$ (parity is not a help here since both parities are present for $K > 0$ and asymmetry splittings are very small).

### E. Coriolis coupling and the in-plane CO bend mode

As shown in Table 3, our calculations agree well with experiment for the ground and first excited vibrational state $K = 1$ energies of both $H_2O$-CO and $D_2O$-CO. In fact, agreement for the excited state (the in-plane CO bend) could be even better than shown since the calculation was based on the ground state intermolecular potential while the experimental values are for the excited intramolecular C-O and O-D stretch states. There is also good agreement with regard to line strengths, with the calculations explaining why the $K = 1 \leftarrow 0$ in-plane CO bend is the only observed combination band and that its line strength is approximately equal to the fundamental $K = 1 \leftarrow 0$ band.

This good agreement gives us confidence that the calculations in Table 1 are reliable, and in particular that the $K = 1$ levels of the intermolecular in-plane CO bend state really do lie below the $K = 0$ levels for both $H_2O$-CO and $D_2O$-CO. This situation ($K = 1$ below $K = 0$, or $\Pi$ state below $\Sigma$ state) is not unprecedented among weakly-bound complexes. It occurs, for example, for *para*-$N_2$ − CO in the excited CO stretch state,[22] and for water dimer in some excited intermolecular vibrations.[23] These examples often involve cases of almost free internal rotation, as for $N_2$ in $N_2$-CO, or else cases of complicated tunneling effects and mixed intermolecular modes, as for water dimer. But neither the water nor the CO motions in the complex are even close to free internal rotations in the present case. Although water-CO is similar in some respects to water dimer, it is still much simpler and more nearly rigid. The tunneling motion in water-CO is significant, but remains a relatively minor effect, especially for $D_2O$-CO. Moreover, the calculations indicate that tunneling is not related to the ordering of $K = 0$ and 1 levels, since the tunneling splittings remain small for the first few excited intermolecular states of $D_2O$-CO.

The wavefunctions show that an 'understanding' (in conventional harmonic mode terms) of water-CO intermolecular vibrations becomes increasingly difficult as vibrational energy increases. But is it still possible to better understand the first few $K = 0$ and 1 levels? In a conventional planar molecule with a large $A$ rotational constant, one of the most important effects for $K > 0$ levels is the $a$-type Coriolis interaction, which links vibrational states of $A'$ and $A''$ symmetry with an interaction strength of $2A\zeta K$, where $\zeta$ is the dimensionless Coriolis coupling parameter whose value can range up to 1. In the present case, we would expect a large

Coriolis coupling between the in-plane and out-of-plane CO bends, with $\zeta$ approaching unity. Indeed, we find that a Coriolis interaction of magnitude $\zeta \approx 0.8$ connecting the in-plane and out-of-plane CO bends would nicely explain the calculated positions of the in-plane CO bend $K = 1$ and 2 levels of $D_2O$-CO at 45 and 62 cm$^{-1}$ (Table 1). It would also work for $K = 1$ of $H_2O$-CO, but not so well for $K = 2$. Of course, this interaction would also push the out-of-plane $K = 1$ and 2 levels up by equal amounts, and this does not agree with the calculated levels. The discrepancy could well be explained by further Coriolis interactions with higher vibrational states which limit the upward push on the out-of-plane $K > 0$ levels. We have not, however, been able to meaningfully extend this Coriolis analysis to these higher vibrational states where the situation rapidly becomes complicated.

Thinking of water-CO as a linear, or quasilinear,[24] molecule provides another way of expressing this same Coriolis mixing. As seen above, the heavy atoms of water-CO are close to being linear near the equilibrium structure. The wavefunctions show that the $\phi$-coordinate "torsional" motion of CO in water-CO is nearly unhindered. When the CO in-plane bend is excited this creates angular momentum, which we can think of as the vibrational angular momentum of an excited bending mode of a quasilinear molecule. In the present case, this bending mode is split into in- and out-of-plane components at about 52 and 70 cm$^{-1}$ for $H_2O$-CO, or 48 and 59 cm$^{-1}$ for $D_2O$-CO. The torsional motion leads to a large internal angular momentum, and the surprising finding that the in-plane CO bend excited $K = 1$ state lies below the corresponding $K = 0$ state can be explained by the large first-order Coriolis coupling between this internal angular momentum and the $a$-axis overall rotation angular momentum.

## V.    Conclusions

Detailed rovibrational energy level calculations, based on a recent high-level intermolecular potential surface,[13] have been made for intermolecular modes of $H_2O$-CO and $D_2O$-CO up to about 120 cm$^{-1}$ above the ground state, for total angular momentum $J = 0$, 1, and 2. In order of increasing energy, the intermolecular modes can be described as: in-plane CO bend (51 cm$^{-1}$), out-of-plane CO bend (70 cm$^{-1}$), van der Waals stretch (78 cm$^{-1}$), in-plane CO bend overtone (89 cm$^{-1}$), with energies as shown for $H_2O$-CO. But above this point, the modes become increasingly mixed and difficult to label. The calculations show that the ground state $K = 1$ levels lie at about 20 cm$^{-1}$ for $H_2O$-CO or 12 cm$^{-1}$ for $D_2$-CO, but that the first excited $K = 1$ levels lie about 3 cm$^{-1}$ *below* their $K = 0$ counterparts. Line strength calculations for infrared bands accompanying intramolecular fundamentals like the C-O stretch and the O-D stretch predict the $K = 0 \leftarrow 0$ fundamental to be the strongest, with the $K = 1 \leftarrow 0$ fundamental being roughly 25 times weaker. But the next $K = 1 \leftarrow 0$ transition to the in-plane CO bend is predicted to have similar strength to the $K = 1 \leftarrow 0$ fundamental, even though the $K = 0 \leftarrow 0$ in-plane CO bend transition is much weaker.

The predicted frequency and line strength of the $K = 1 \leftarrow 0$ transition to the in-plane CO bend is strikingly verified for $H_2O$-CO and $D_2$-CO by new mid-infrared spectroscopic measurements. These experiments are performed using a pulsed supersonic slit jet expansion which is probed by a tunable infrared quantum cascade laser or optical parametric oscillator source. They cover the C-O stretch region around 2200 cm$^{-1}$ for $H_2O$-CO, $D_2O$-CO, and HOD-CO, as well as the O-D stretch region around 2700 cm$^{-1}$ for $D_2O$-CO, HOD-CO, and DOH-CO. Observations of discrete perturbations in the symmetric O-D stretch region enable an experimental lower limit of about 340 cm$^{-1}$ to be established for $D_2O$-CO, to be compared with our calculated binding energy of 368 cm$^{-1}$. Wavefunction plots for various states of H$_2$O-CO are examined to help understand its intermolecular dynamics, and significant Coriolis interactions are invoked to explain the seemingly anomalous energies of the first excited $K = 0$ and 1 levels of H$_2$O-CO and D$_2$O-CO.

**Supplementary Material**

Supplementary Material includes tables giving calculated rovibrational levels for H$_2$O-CO and D$_2$O-CO.

**Acknowledgements**

The financial support of the Natural Sciences and Engineering Research Council of Canada is gratefully acknowledged. We thank K. Michaelian for the loan of the QCL.

Table 1. Calculated intermolecular vibrational levels of $H_2O$-CO and $D_2O$-CO.[a]

| Vibrational mode | $K$ | H2O-CO | | | D2O-CO | | |
|---|---|---|---|---|---|---|---|
| | | A | B | Line str.[b] | A | B | Line str.[b] |
| ground state | 0e | 0.000 | 0.795 | 1.0 | 0.000 | 0.030 | 1.0 |
| ground state | 1 | 19.868 | 19.666 | 0.0498 | 11.618 | 11.598 | 0.0332 |
| ground state | 2 | 57.327 | 57.456 | | 44.033 | 44.024 | |
| i-p CO bend | 0e | 51.270 | 52.501 | 0.0010 | 48.409 | 48.528 | 0.0009 |
| i-p CO bend | 1 | 49.429 | 48.544 | 0.0444 | 45.432 | 45.349 | 0.0530 |
| | 2 | 86.619 | 85.830 | | 61.638 | 61.726 | |
| o-p CO bend | 0f | 70.704 | 68.998 | 0.0 | 59.006 | 58.915 | 0.0 |
| | 1 | 80.551 | 81.421 | 0.0033 | 75.778 | 75.814 | 0.0034 |
| | 2 | 103.017 | 104.640 | | 89.233 | 89.370 | |
| vdW stretch | 0e | 77.807 | 79.073 | 0.0005 | 77.075 | 77.130 | <0.0001 |
| vdW stretch | 1 | 96.531 | 95.065 | <0.0001 | 89.249 | 89.005 | <0.0001 |
| i-p CO bend overtone? | 0e | 88.419 | 89.710 | 0.0002 | 84.008 | 84.340 | 0.0005 |
| | 1 | 108.259 | 107.160 | 0.0001 | 95.223 | 95.172 | 0.0002 |
| i-p CO bend + vdW stretch? | 0e | 111.388 | 112.935 | 0.0003 | 110.050 | 110.237 | |
| | 1 | 123.05 | 123.27 | | 104.260 | 104.379 | <0.0001 |
| i-p + o-p CO bend [c] | 0f | ? | 124.875 | | 114.468 | 114.265 | |

[a] Calculated origins are given for $K = 1$ (or 2), defined as in Eq. 1. These have 1*B (or 2*B) subtracted from actual calculated $J = 1$ (or 2) levels, using B = 0.092 and 0.087 cm$^{-1}$ for H2O and D2O-CO, respectively. Only the lowest three $K = 2$ states are shown. i-p = in-plane; o-p = out-of-plane; vdW = van der Waals.
[b] Calculated C-O stretch region relative line strengths for $R(0)$ transitions originating in the ground $K = 0$ state.
[c] Or o-p $H_2O$ bend?

Table 2. Fitted spectroscopic parameters for $H_2O$-CO (in cm$^{-1}$).

| | $K$ | $\sigma$ | $(B + C)/2$ | $10^4 \times (B - C)$ | $10^7 \times D_J$ |
|---|---|---|---|---|---|
| A, Ground state [a] | 0 | 0.0 | [0.09170104] | | [6.829] |
| B, Ground state [a] | 0 | 0.0 [c] | [0.09174705] | | [6.810] |
| A, Ground state [a] | 1 | [19.8337301] | [0.09254137] | [8.7107] | [8.157] |
| B, Ground state [a] | 1 | [18.7207181] | [0.09242805] | [6.3175] | [8.090] |
| A, C-O stretch | 0 | 2153.5942(1) | 0.0912423(59) | | 6.37(86) |
| B, C-O stretch | 0 | 2153.6470(1) | 0.0912998(45) | | 7.42(53) |
| A, C-O stretch | 1 | 2173.3348(1) | 0.092156(26) | 8.71(18) | 6.7(48) |
| B, C-O stretch | 1 | 2172.2414(1) | 0.092109(11) | 8.72(12) | 11.6(14) |
| A, C-O stretch combination | 1 | 2201.4490(2) | 0.092487(24) | 25.32(16) | 1.9(41) |
| B, C-O stretch combination | 1 | 2199.4372(2) | 0.092895(15) | 7.55(12) | 11.4(18) |

[a] Ground state parameters from Bumgarner et al.[3] were fixed in the infrared fits. Additional ground state parameters are, for the A state: $\delta_J = 1.73 \times 10^{-8}$ cm$^{-1}$; for the B state: $\delta_J = 3.85 \times 10^{-8}$, $h_J = 4.20 \times 10^{-11}$ cm$^{-1}$. The $D_{JK}$ and $H_{JK}$ parameters from Ref. 3 are incorporated into the quoted $(B + C)/2$ and $D_J$ values for $K = 1$.

Table 3. Calculated and observed origins of $K = 1$ states of the fundamental and intermolecular in-plane CO bend vibrations, relative to the respective A or B state fundamental $K = 0$ origin (in cm$^{-1}$).[a]

|  |  | Ground state theory | Ground state experiment [3] | C-O stretch experiment | O-D stretch-3 experiment [b] |
|---|---|---|---|---|---|
| Fundamental $K = 1$ | H$_2$O-CO | 19.868, 18.871 | 19.834, 18.721 | 19.741, 18.594 |  |
|  | D$_2$O-CO | 11.618, 11.568 | 11.784, 11.716 | 11.755, 11.702 | 11.386, 11.450 |
|  | HOD-CO |  |  | 18.966 |  |
| In-plane CO bend $K = 1$ | H$_2$O-CO | 49.429, 47.749 |  | 47.855, 45.790 |  |
|  | D$_2$O-CO | 45.432, 45.319 |  | 43.638, 43.482 | 44.737, 44.825 |
|  | HOD-CO |  |  | 48.096 |  |

[a] The two values given for H$_2$O-CO and D$_2$O-CO are for the A and B tunneling states.

[b] For D$_2$O-CO in the O-D stretch-1 state, we obtain an approximate value of 11.62 cm$^{-1}$ for the fundamental $K = 1$ energy, which is an average for the A and B states.

Table 4. Fitted spectroscopic parameters for $D_2O$-CO (in $cm^{-1}$).

| | $K$ | $\sigma$ | $(B+C)/2$ | $10^4 \times (B-C)$ | $10^7 \times D_J$ |
|---|---|---|---|---|---|
| A, Ground state [a] | 0 | 0.0 | [0.08736800] | | [6.64] |
| B, Ground state [a] | 0 | 0.0 [c] | [0.08735768] | | [4.822] |
| A, Ground state [a] | 1 | [11.7837644] | [0.08747577] | [8.5953] | [8.91] |
| B, Ground state [a] | 1 | [11.7161897] | [0.08747506] | [8.3641] | [8.52] |
| A, C-O stretch | 0 | 2154.5369(1) | 0.0869374(48) | | 6.81(56) |
| B, C-O stretch | 0 | 2154.5404(1) | 0.0869184(48) | | 3.98(56) |
| A, C-O stretch | 1 | 2166.3117(1) | 0.0869911(39) | 6.62(15) | 6.25(89) |
| B, C-O stretch | 1 | 2166.2426(1) | 0.0869904(83) | 5.85(30) | 5.7(28) |
| A, C-O stretch combination | 1 | 2198.1745(1) | 0.087726(13) | 10.601(78) | 1.7(21) |
| B, C-O stretch combination | 1 | 2198.0219(1) | 0.087895(15) | 9.02(10) | 6.3(28) |
| A, O-D stretch-3 | 0 | 2780.6776(1) | 0.0872707(70) | | 4.56(83) |
| B, O-D stretch-3 | 0 | 2780.5988(1) | 0.0873298(70) | | 8.45(83) |
| A, O-D stretch-3 | 1 | 2792.0631(1) | 0.0873287(44) | 6.64(17) | 6.03(97) |
| B, O-D stretch-3 | 1 | 2792.0488(2) | 0.0873320(61) | 8.00(23) | 3.9(16) |
| A, O-D stretch-3 combination | 1 | 2825.4141(1) | 0.089400(13) | 10.23(11) | 7.6(24) |
| B, O-D stretch-3 combination | 1 | 2825.4238(1) | 0.089720(14) | 12.73(16) | 26.4(27) |
| O-D stretch-1 [b] | 0 | 2665.2356(5) | 0.0872589(39) | | -9.5(57) |
| A, O-D stretch-1 | 1 | 2676.885(3) | 0.08732(14) | [7.0] | [6.0] |
| B, O-D stretch-1 | 1 | 2676.836(5) | 0.08750(28) | [7.0] | [6.0] |

[a] Ground state parameters from Bumgarner et al.[3] were fixed in the infrared fits. Additional ground state parameters are, for the A state: $H_J = 7.9 \times 10^{-12}$, $h_{JK} = 1.9 \times 10^{-7}$ $cm^{-1}$; for the B state: $H_J = 1.59 \times 10^{-12}$ $cm^{-1}$. The $D_{JK}$ and $H_{JK}$ parameters from Ref. 3 are incorporated into the quoted $(B+C)/2$ and $D_J$ values for $K = 1$.

[b] For O-D stretch-1, the $K = 0$ fit used the centers of gravity of the lines (see text), and should represent a deperturbed average of A and B components. The O-D stretch-1 $K = 1$ fits are very approximate.

Table 5. Fitted spectroscopic parameters for HOD-CO and DOH-CO (in cm$^{-1}$).

| | $K$ | $\sigma$ | $(B + C)/2$ | $(B - C)$ | $D_J$ |
|---|---|---|---|---|---|
| HOD-CO ground state [a] | 0 | 0.0 | [0.09097574] | | [6.04 × 10$^{-7}$] |
| HOD-CO C-O stretch | 0 | 2154.4795(1) | 0.0905204(49) | | 7.40(69) × 10$^{-7}$ |
| HOD-CO C-O stretch | 1 | 2173.4453(1) | 0.0911901(22) | 6.928(87) × 10$^{-4}$ | [7.40 × 10$^{-7}$] |
| HOD-CO C-O stretch combination | 1 | 2202.5752(1) | 0.0916941(21) | 9.632(51) × 10$^{-4}$ | [7.40 × 10$^{-7}$] |
| HOD-CO O-D stretch | 0 | 2709.4788(1) | 0.0908641(33) | | 4.74(27) × 10$^{-7}$ |
| HOD-$^{13}$CO O-D stretch [a] | 0 | 2709.4236(1) | 0.0904270(42) | | [6.04 × 10$^{-7}$] |
| DOH-CO ground state | 0 | 0.0 | 0.088013(13) | | 5.6(26) × 10$^{-7}$ |
| DOH-CO C-O stretch | 0 | 2153.7413(1) | 0.087590(13) | | 5.4(21) × 10$^{-7}$ |
| DOH-CO O-D stretch | 0 | 2725.4004(1) | 0.087948(13) | | 2.3(21) × 10$^{-7}$ |

[a] HOD-CO ground state parameters were fixed at these values from Yaron et al.[2] Not shown are those for HOD-$^{13}$CO ground state, $(B + C)/2 = 0.09052009$, $D = 6.04 \times 10^{-7}$ cm-1. Parameters in square brackets were fixed in the fits.

**Figure Captions**

Fig. 1. Planar equilibrium structure of $H_2O$-CO and with the coordinates $R$ = 3.93 Å = 7.42 $a_0$, $\theta_1$ = 115.3°, $\theta_2$ = 8.6°, $\phi$ = 0°, and $\chi$ = 0°. The definition of these coordinates with respect to a dimer-fixed frame with its z-axis along the intermolecular axis **R** is given in the text.

Fig. 2. Spectra of $H_2O$-CO (left) and $D_2$-CO (right) in the C-O stretch region. Upper panels show the $K$ = 1 ← 0 fundamental band, and lower panels show the $K$ = 1 ← 0 combination band involving the intermolecular in-plane CO bending mode. Simulated spectra illustrate the contributions of the A (blue) and B (red) tunneling components. The $K$ = 0 ← 0 fundamental bands have been studied previously.[1]

Fig. 3. Spectra of $D_2O$-CO in the O-D stretch-3 region. The lower panel shows the fundamental $K$ = 0 ← 0 band, the middle panel shows the fundamental $K$ = 1← 0 band, and the upper panel shows the combination $K$ = 1← 0 band. Simulated spectra illustrate the contributions of the A (blue) and B (red) tunneling components.

Fig. 4. Spectra of $D_2O$-CO in the O-D stretch-1 region. The lower panel shows the fundamental $K$ = 0 ← 0 band, and the upper panel shows the fundamental $K$ = 1← 0 band. Both bands are perturbed, such that the A and B tunneling components cannot be distinguished in the $K$ = 0 ← 0 band.

Fig. 5. Spectra of HOD-CO and DOH-CO. The left-hand panels show the HOD-CO $K$ = 1 ← 0 fundamental (bottom) and $K$ = 1 ← 0 combination bands (top) in the C-O stretch region. The right-hand panels show the HOD-CO (bottom) and DOH-CO (top) $K$ = 0 ← 0 fundamental bands in the O-D stretch region. Simulated spectra are in red (there is no tunneling splitting for these species).

Fig. 6. Wavefunction contour plots for the ground $K = 0$ state of $H_2O$-CO. The four upper panels are for the A tunneling component, and the four lower panels for the B component. The only significant difference between A and B lies in the relative phase of the two nodes in the $\theta(H_2O)$ vs. $\chi(H_2O)$ plots.

Fig. 7. Wavefunction contour plots for the first excited $K = 0$ (upper four panels) and $K = 1$ (lower four panels) state of $H_2O$-CO. This state is assigned as the intermolecular in-plane CO bend.

Fig. 8. Wavefunction contour plots for the second (upper four panels) and third (lower four panels) excited $K = 0$ states of $H_2O$-CO. These are assigned as the intermolecular out-of-plane CO bend and van der Waals stretch modes, respectively.

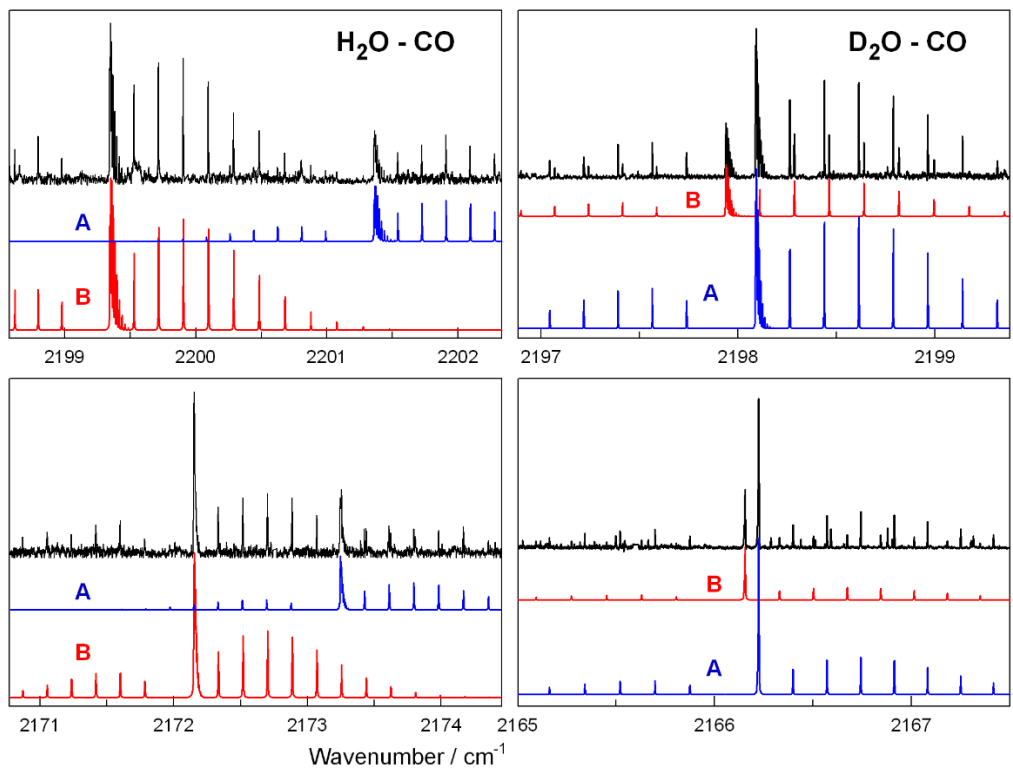

H₂O - CO

A

B

D₂O - CO

B

A

A

B

B

A

Wavenumber / cm⁻¹

Fig. 2.

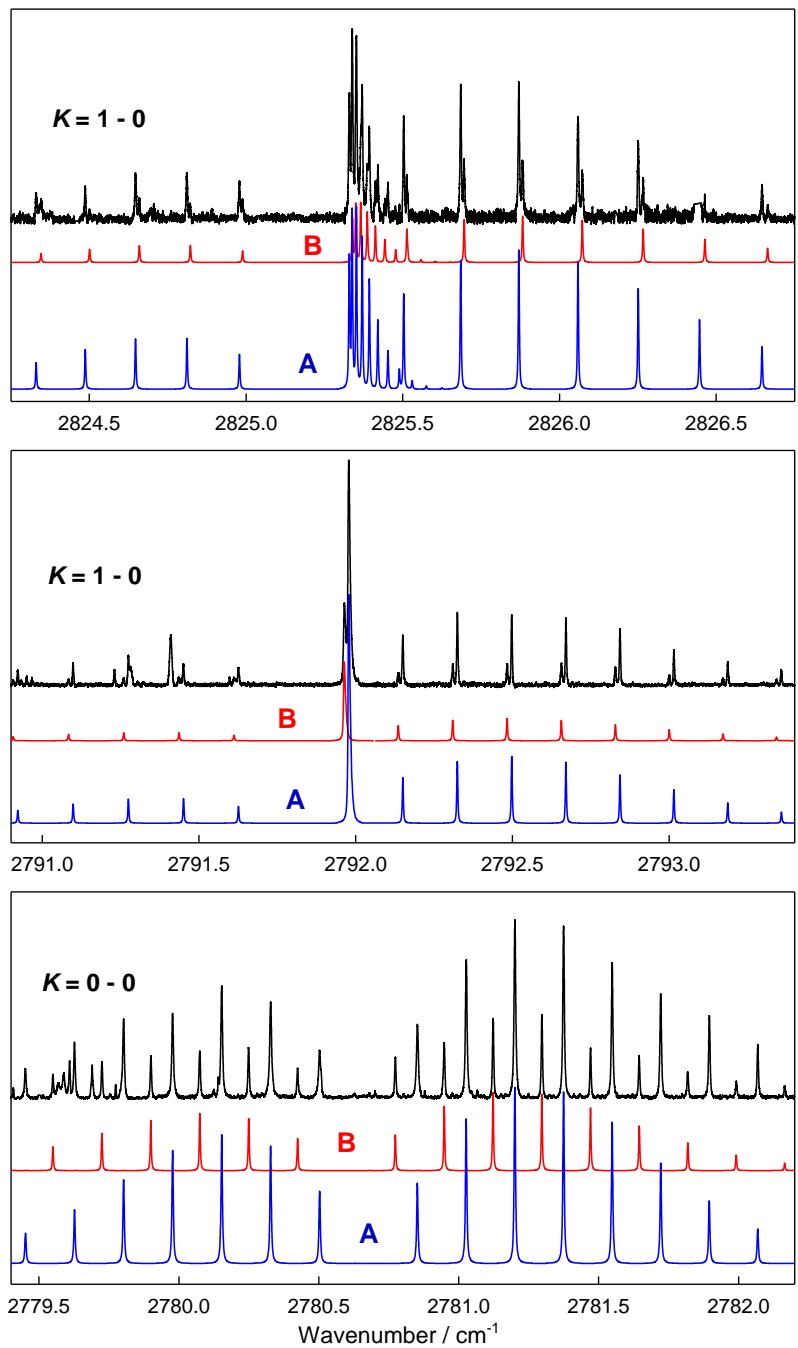

Fig. 3.

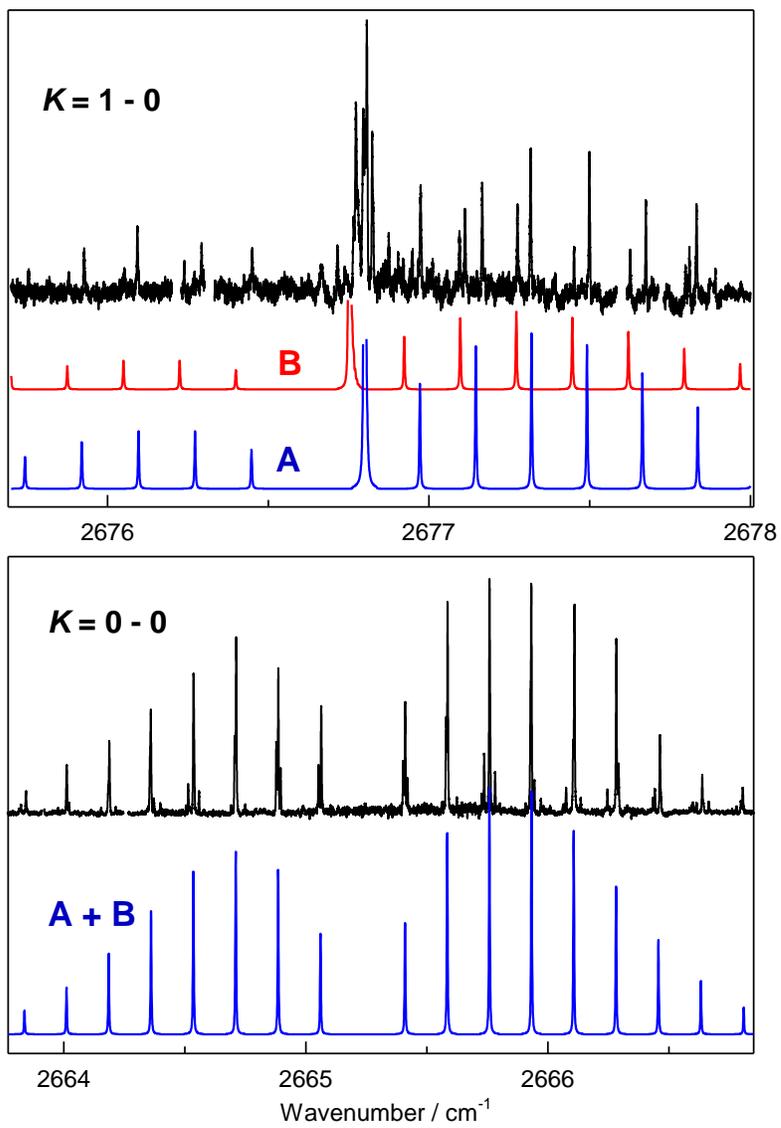

Fig. 4.

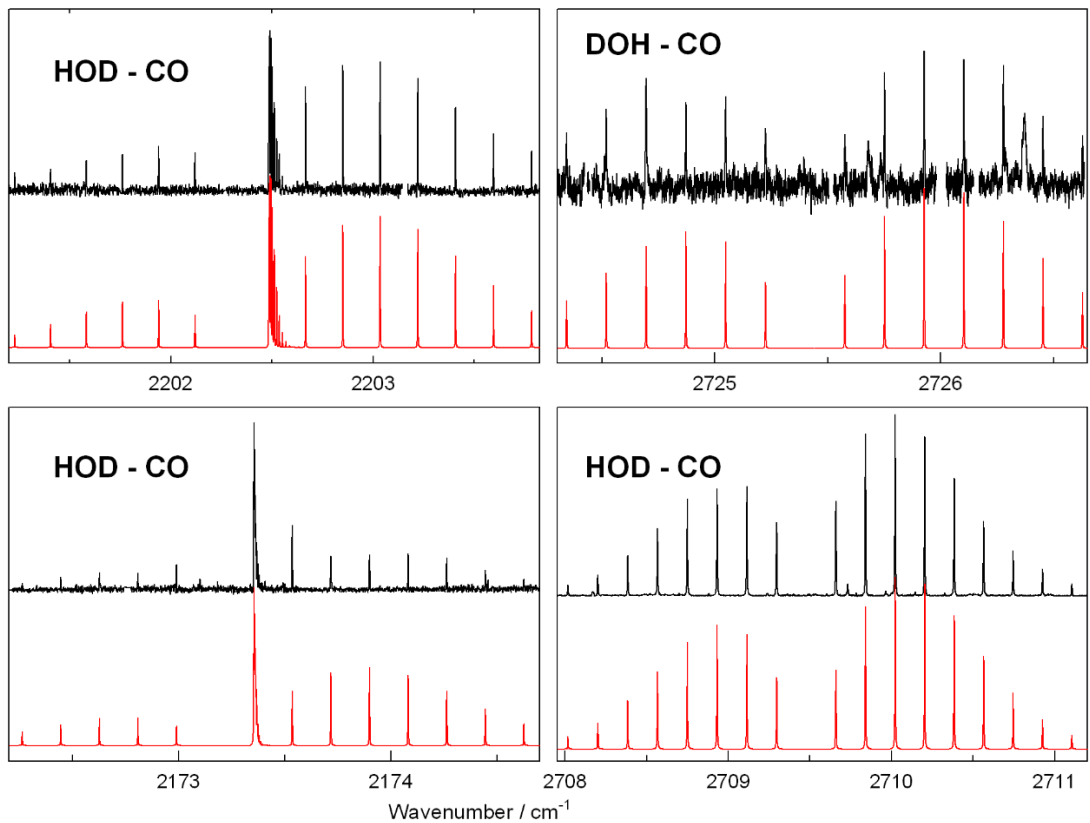

Fig. 5.

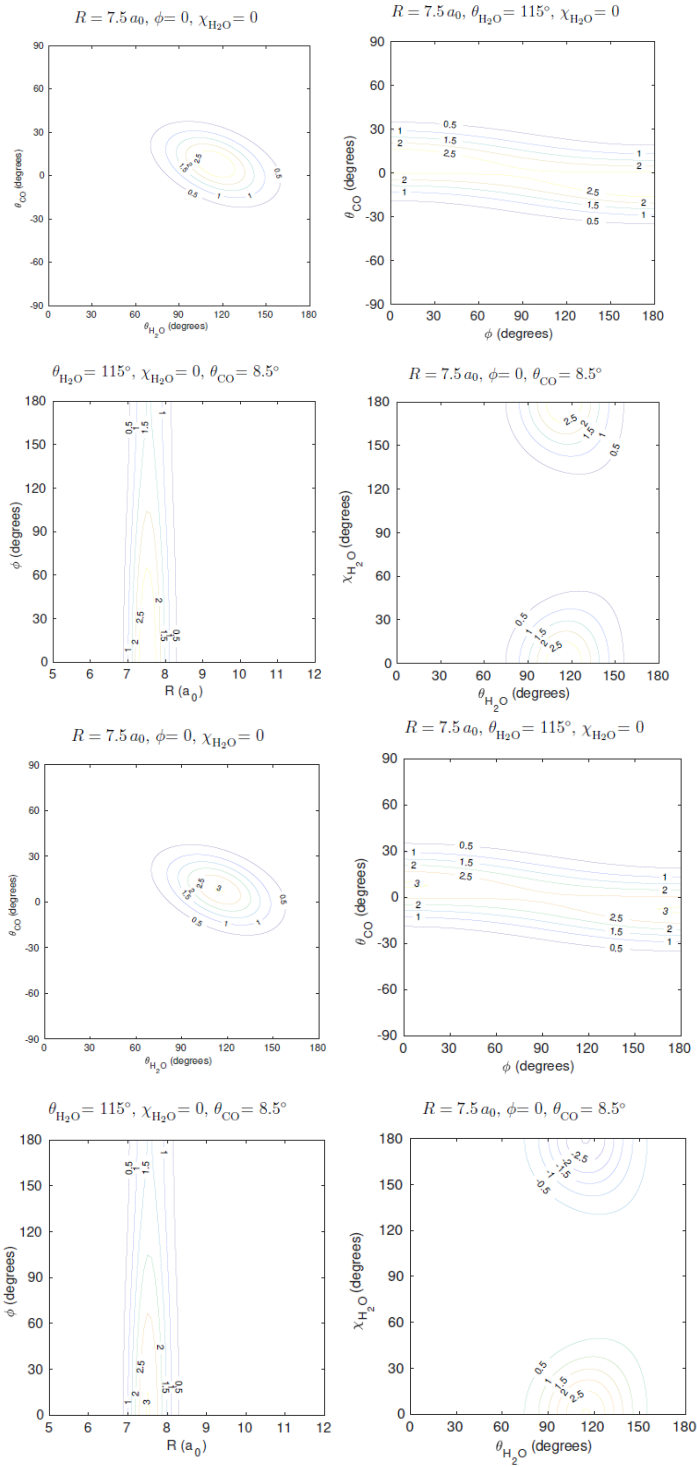

Fig. 6

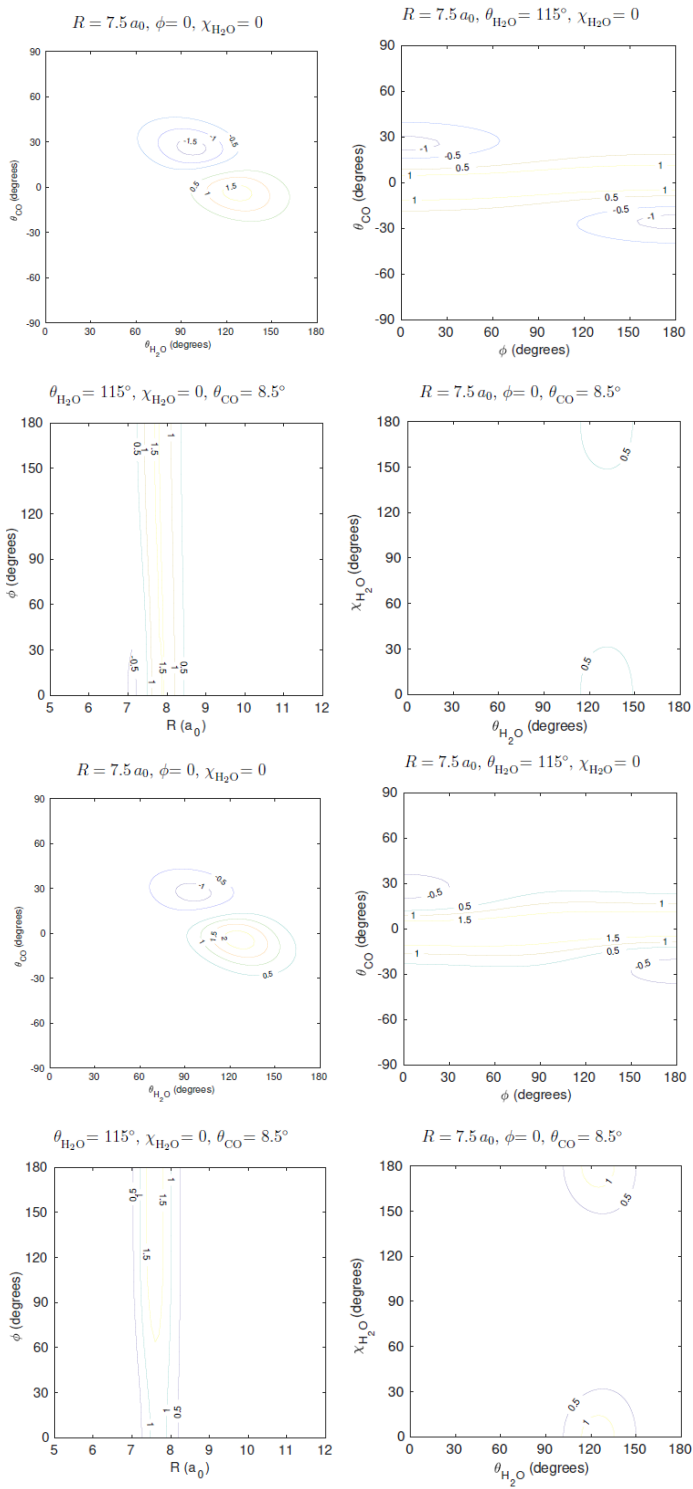

Fig. 7.

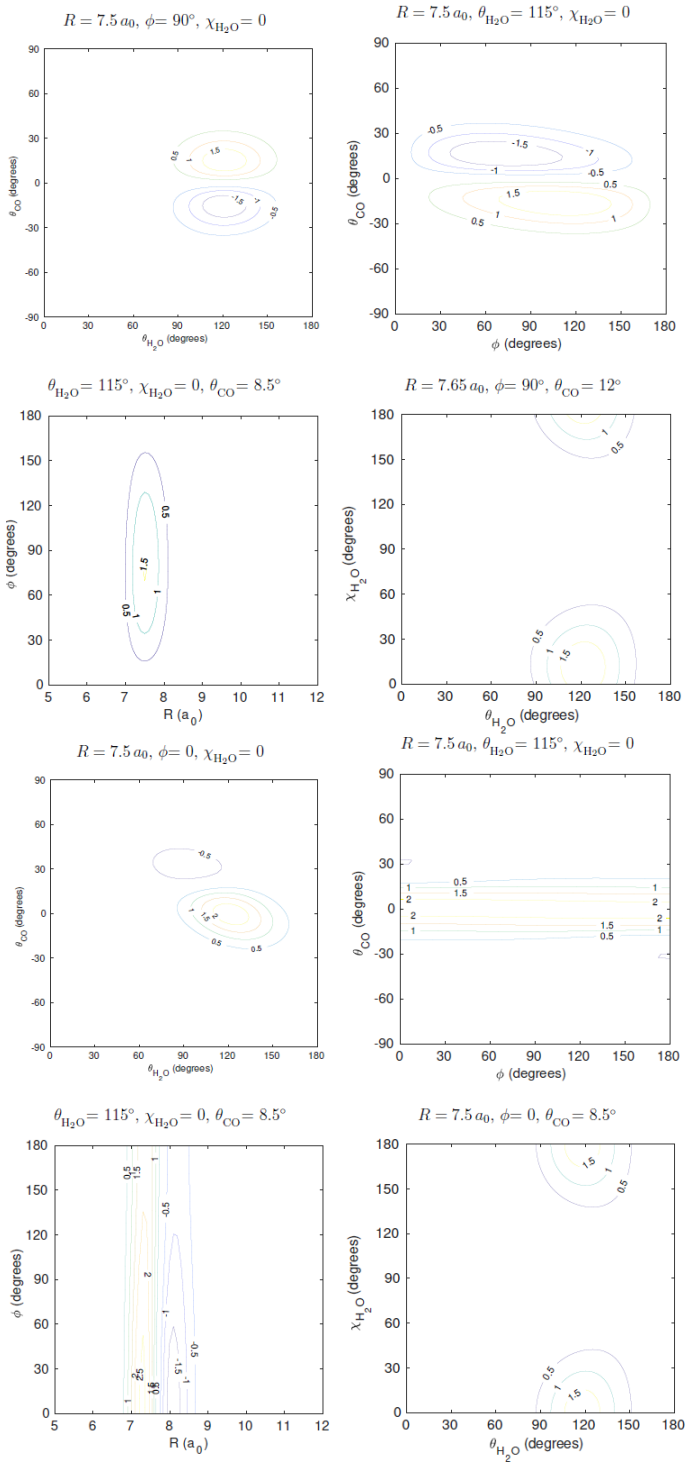

Fig. 8.

Supplementary Information for:

**The water – carbon monoxide dimer: new infrared spectra, *ab initio* rovibrational energy level calculations, and an interesting intermolecular mode**

By:

A. Barclay,[†] A. van der Avoird,[‡] A.R.W. McKellar,[§] and N. Moazzen-Ahmadi

[†]*Department of Physics and Astronomy, University of Calgary, 2500 University Drive North West, Calgary, Alberta T2N 1N4, Canada,* [‡]*Theoretical Chemistry, Institute for Molecules and Materials, Radboud University, Heyendaalseweg 135,6525 AJ Nijmegen, The Netherlands,* [§]*National Research Council of Canada, Ottawa, Ontario K1A 0R6, Canada.*

Table A-1. Calculated energy levels for the A state of CO-H$_2$O (*para* H$_2$O – CO), relative to dissociation (columns on left) and relative to the lowest level (columns on right) (in cm$^{-1}$).

| K | J = 0 | J = 1 | J = 2 | J = 0 | J = 1 | J = 2 |
|---|-------|-------|-------|-------|-------|-------|
| Parity = e | | | | | | |
| 0 | -315.9595 | -315.7764 | -315.4101 | 0.0000 | 0.1831 | 0.5494 |
| 1 | | -295.9991 | -295.6306 | | 19.9604 | 20.3289 |
| 1 | | -266.4385 | -266.0694 | | 49.5210 | 49.8901 |
| 0 | -264.6893 | -264.4997 | -264.1205 | 51.2702 | 51.4598 | 51.8390 |
| 2 | | | -258.4495 | | | 57.5100 |
| 0 | -238.1530 | -237.9762 | -237.6227 | 77.8065 | 77.9833 | 78.3368 |
| 1 | | -235.3142 | -234.9368 | | 80.6453 | 81.0227 |
| 2 | | | -229.1576 | | | 86.8019 |
| 0 | -227.5408 | -227.3511 | -226.9716 | 88.4187 | 88.6084 | 88.9879 |
| 1 | | -219.3368 | -218.9698 | | 96.6227 | 96.9897 |
| 2 | | | -212.7598 | | | 103.1997 |
| 1 | | -207.6071 | -207.2321 | | 108.3524 | 108.7274 |
| 0 | -204.5711 | -204.3823 | -204.0045 | 111.3884 | 111.5772 | 111.9550 |
| 2 | | | -198.8401 | | | 117.1194 |
| Parity = f | | | | | | |
| 1 | | -295.9983 | -295.6281 | | 19.9612 | 20.3314 |
| 1 | | -266.4369 | -266.0645 | | 49.5226 | 49.8950 |
| 2 | | | -258.4495 | | | 57.5100 |
| 0 | -245.2553 | -245.0701 | -244.6999 | 70.7042 | 70.8894 | 71.2596 |
| 1 | | -235.3127 | -234.9324 | | 80.6468 | 81.0271 |
| 2 | | | -229.1576 | | | 86.8019 |
| 1 | | -219.3355 | -218.9660 | | 96.6240 | 96.9935 |
| 2 | | | -212.7598 | | | 103.1997 |
| 1 | | -207.6069 | -207.2317 | | 108.3526 | 108.7278 |
| 2 | | | -198.8401 | | | 117.1194 |
| 1 | | -192.8147 | -192.4286 | | 123.1448 | 123.5309 |

Table A-2. Calculated energy levels for the B state of CO-H$_2$O (*ortho* H$_2$O – CO), relative to A state dissociation (columns on left) and relative to the lowest A state level (columns on right) (in cm$^{-1}$). (The B state dissociation threshold is +23.2994 cm$^{-1}$).

| K | J = 0 | J = 1 | J = 2 | J = 0 | J = 1 | J = 2 |
|---|---|---|---|---|---|---|
| Parity = *e* | | | | | | |
| 0 | -315.1644 | -314.9812 | -314.6148 | 0.7951 | 0.9783 | 1.3447 |
| 1 | | -296.2015 | -295.8332 | | 19.7580 | 20.1263 |
| 1 | | -267.3232 | -266.9523 | | 48.6363 | 49.0072 |
| 0 | -263.4585 | -263.2690 | -262.8902 | 52.5010 | 52.6905 | 53.0693 |
| 2 | | | -258.3207 | | | 57.6388 |
| 0 | -236.8866 | -236.7101 | -236.3571 | 79.0729 | 79.2494 | 79.6024 |
| 1 | | -234.4441 | -234.0663 | | 81.5154 | 81.8932 |
| 2 | | | -229.9466 | | | 86.0129 |
| 0 | -226.2499 | -226.0598 | -225.6797 | 89.7096 | 89.8997 | 90.2798 |
| 1 | | -220.8027 | -220.4364 | | 95.1568 | 95.5231 |
| 2 | | | -211.1363 | | | 104.8232 |
| 1 | | -208.7053 | -208.3292 | | 107.2542 | 107.6303 |
| 0 | -203.0244 | -202.8357 | -202.4582 | 112.9351 | 113.1238 | 113.5013 |
| 2 | | | -198.5718 | | | 117.3877 |
| Parity = *f* | | | | | | |
| 1 | | -296.2009 | -295.8314 | | 19.7586 | 20.1281 |
| 1 | | -267.3224 | -266.9502 | | 48.6371 | 49.0093 |
| 2 | | | -258.3207 | | | 57.6388 |
| 0 | -246.9614 | -246.7761 | -246.4054 | 68.9981 | 69.1834 | 69.5541 |
| 1 | | -234.4430 | -234.0628 | | 81.5165 | 81.8967 |
| 2 | | | -229.9466 | | | 86.0129 |
| 1 | | -220.8019 | -220.4339 | | 95.1576 | 95.5256 |
| 2 | | | -211.1363 | | | 104.8232 |
| 1 | | -208.7061 | -208.3315 | | 107.2534 | 107.6280 |
| 2 | | | -198.5718 | | | 117.3877 |
| 1 | | -192.5955 | -192.2296 | | 123.3640 | 123.7299 |
| 0 | -191.0843 | -190.8949 | -190.5162 | 124.8752 | 125.0646 | 125.4433 |
| 1 | | -190.7663 | -190.3866 | | 125.1932 | 125.5729 |
| 2 | | | -185.1680 | | | 130.7915 |
| 1 | | -176.2549 | -175.8844 | | 139.7046 | 140.0751 |

Table A-3. Calculated energy levels for the A state of CO-D$_2$O (*ortho* D$_2$O – CO), relative to dissociation (columns on left) and relative to the lowest level (columns on right) (in cm$^{-1}$).

| K | J = 0 | J = 1 | J = 2 | J = 0 | J = 1 | J = 2 |
|---|---|---|---|---|---|---|
| Parity = *e* | | | | | | |
| 0 | -368.4166 | -368.2432 | -367.8963 | 0.0000 | 0.1734 | 0.5203 |
| 1 | | -356.7124 | -356.3660 | | 11.7042 | 12.0506 |
| 2 | | | -324.2084 | | | 44.2082 |
| 1 | | -322.8967 | -322.5442 | | 45.5199 | 45.8724 |
| 0 | -320.0076 | -319.8294 | -319.4731 | 48.4090 | 48.5872 | 48.9435 |
| 2 | | | -306.6043 | | | 61.8123 |
| 1 | | -292.5510 | -292.1993 | | 75.8656 | 76.2173 |
| 0 | -291.3413 | -291.1732 | -290.8370 | 77.0753 | 77.2434 | 77.5796 |
| 0 | -284.4091 | -284.2302 | -283.8723 | 84.0075 | 84.1864 | 84.5443 |
| 2 | | | -279.0090 | | | 89.4076 |
| 1 | | -279.077 | -278.7128 | | 89.3396 | 89.7038 |
| 1 | | -273.1076 | -272.7640 | | 95.3090 | 95.6526 |
| 1 | | -264.0684 | -263.7139 | | 104.3482 | 104.7027 |
| 2 | | | -259.5412 | | | 108.8754 |
| Parity = *f* | | | | | | |
| 1 | | -356.7116 | -356.3634 | | 11.7050 | 12.0532 |
| 2 | | | -324.2084 | | | 44.2082 |
| 1 | | -322.8957 | -322.5410 | | 45.5209 | 45.8756 |
| 0 | -309.4110 | -309.2353 | -308.8841 | 59.0056 | 59.1813 | 59.5325 |
| 2 | | | -306.6043 | | | 61.8123 |
| 1 | | -292.5501 | -292.1966 | | 75.8665 | 76.2200 |
| 2 | | | -279.0089 | | | 89.4077 |
| 1 | | -279.0758 | -278.7093 | | 89.3408 | 89.7073 |
| 1 | | -273.1073 | -272.7634 | | 95.3093 | 95.6532 |
| 1 | | -264.0682 | -263.7133 | | 104.3484 | 104.7033 |
| 2 | | | -259.5412 | | | 108.8754 |
| 0 | -253.9484 | -253.7691 | -253.4105 | 114.4682 | 114.6475 | 115.0061 |
| 2 | | | -247.3447 | | | 121.0719 |
| 1 | | -247.4532 | -247.0963 | | 120.9634 | 121.3203 |
| 1 | | -244.9276 | -244.5752 | | 123.4890 | 123.8414 |

Table A-4. Calculated energy levels for the B state of CO-D₂O (*para* D₂O − CO), relative to A state dissociation (columns on left) and relative to the lowest A state level (columns on right) (in cm⁻¹). (The B state dissociation threshold is +12.1183 cm⁻¹).

| $K$ | $J = 0$ | $J = 1$ | $J = 2$ | $J = 0$ | $J = 1$ | $J = 2$ |
|-----|---------|---------|---------|---------|---------|---------|
| Parity = *e* | | | | | | |
| 0 | -368.3866 | -368.2132 | -367.8663 | 0.03000 | 0.2034 | 0.5503 |
| 1 | | -356.7317 | -356.3852 | | 11.6849 | 12.0314 |
| 2 | | | -324.2174 | | | 44.1992 |
| 1 | | -322.9799 | -322.6270 | | 45.4367 | 45.7896 |
| 0 | -319.889 | -319.7109 | -319.3548 | 48.5276 | 48.7057 | 49.0618 |
| 2 | | | -306.5157 | | | 61.9009 |
| 1 | | -292.515 | -292.1633 | | 75.9016 | 76.2533 |
| 0 | -291.2868 | -291.1188 | -290.7828 | 77.1298 | 77.2978 | 77.6338 |
| 0 | -284.0765 | -283.8975 | -283.5395 | 84.3401 | 84.5191 | 84.8771 |
| 2/1 | | | -279.0124 | | | 89.4042 |
| 1/2 | | -279.3097 | -278.9016 | | 89.1069 | 89.5150 |
| 1 | | -273.1584 | -272.8147 | | 95.2582 | 95.6019 |
| 1 | | -263.9491 | -263.5945 | | 104.4675 | 104.8221 |
| 2 | | | -259.5414 | | | 108.8752 |
| Parity = *f* | | | | | | |
| 1 | | -356.7309 | -356.3827 | | 11.6857 | 12.0339 |
| 2 | | | -324.2174 | | | 44.1992 |
| 1 | | -322.9799 | -322.6243 | | 45.4376 | 45.7923 |
| 0 | -309.5019 | -309.3262 | -308.9748 | 58.9147 | 59.0904 | 59.4418 |
| 2 | | | -306.5157 | | | 61.9009 |
| 1 | | -292.5142 | -292.1608 | | 75.9024 | 76.2558 |
| 2/1 | | | -279.0110 | | | 89.4056 |
| 1/2 | | -279.3086 | -278.9000 | | 89.1080 | 89.5166 |
| 1 | | -273.1582 | -272.8144 | | 95.2584 | 95.6022 |
| 1 | | -263.9493 | -263.5952 | | 104.4673 | 104.8214 |
| 2 | | | -259.5414 | | | 108.8752 |
| 0 | -254.1514 | -253.9719 | -253.6130 | 114.2652 | 114.4447 | 114.8036 |
| 2 | | | -247.5473 | | | 120.8693 |
| 1 | | -247.6461 | -247.2889 | | 120.7705 | 121.1277 |
| 1 | | -245.2796 | -244.9290 | | 123.1370 | 123.4876 |